# Designing a User Interface for Generative Design in Augmented Reality: A Step Towards More Visualization and Feed-Forwarding

| Sora Kang | Kaiwen Yu | Xinyi Zhou | Joonhwan Lee |
|---|---|---|---|
| HCI+D lab, Seoul National University | University of California, Berkeley | University of California, Berkeley | HCI+D lab, Seoul National University |
| sorakang@snu.ac.kr | kaiwen_yu@berkely.edu | xinyi_zhou0819@berkeley.edu | joonhwan@snu.ac.kr |

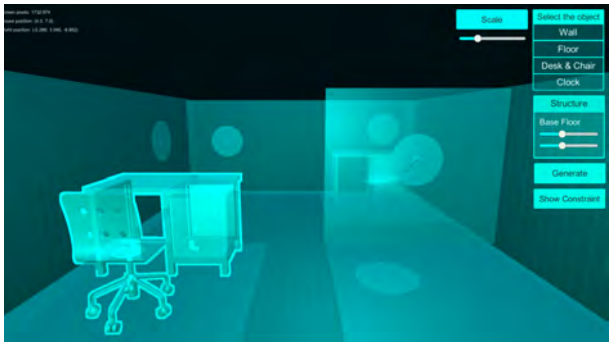

Figure 1 Demonstration of the UI System in Action

## Abstract

Generative design, an AI-assisted technology for optimizing design through algorithmic processes, is propelling advancements across numerous fields. As the use of immersive environments such as Augmented Reality (AR) continues to rise, integrating generative design into such platforms presents a potent opportunity for innovation. However, a vital challenge that impedes this integration is the current absence of an efficient and user-friendly interface for designers to operate within these environments effectively. To bridge this gap, we introduce a novel UI system for generative design software in AR, which automates the process of generating the potential design constraints based on the users' inputs. This system allows users to construct a virtual environment, edit objects and constraints, and export the final data in CSV format. The interface enhances the user's design experience by enabling more intuitive interactions and providing immediate visual feedback. Deriving from participatory design principles, this research proposes a significant leap forward in the realms of generative design and immersive environments.

## 1. Introduction

Generative design, a prominent concept in architecture and civil engineering, has now made its distinguished footprint in mechanical design. Recognized as a revolutionary approach, generative design optimally assists designers in their creative activities. As stated by CIMdata, generative design entails an interactive approach that utilizes algorithmic principles to morph necessities and limitations into a final product design, a process that goes beyond physics and organic methodologies. The design refinement involves repeated iterations and tweaking constraints as per the output. However, conventional product development is often constrained by team capacity, time, money, and a limited range of iterations, causing the end product deprived of optimal solutions. Enter generative design. The approach presents an array of alternative solutions with various degrees of departure from the original outline in an expedited manner.

Over the years, industries such as automotive, aerospace, and sports have reaped the benefits of generative design. The method has also marked its presence in fashion and furniture design, with the launch of the first commercial chair utilizing generative design back in 2019. However, the implementation of generative design in immersive environments such as augmented reality (AR) remains scant. Simultaneously, the market for AR has seen a significant influx of large tech companies vying for its potential – one that reaches beyond gaming to industries such as automobile sales and corporate training. Predicted sales for AR and VR headsets could hit $80 million by 2021. Despite these prospects, there is an obvious void – a generative design software compatible with the immersive environment of AR and VR.

The transition of generative design into the AR environment brings several benefits. Digitally, we can place objects in physical environments – a necessity for AR experiences. Keeping this in mind, an AR application should adapt to the myriad of environments it is used in – the workplace, home, or natural surroundings. This necessitates a generative approach when designing the layout of the digital objects – defining constraints and pre-conditions at a higher level abstraction. This is akin to the concept of generative design – constraints are fixed and ample candidate designs are generated according to these constraints. However, the results are divergent in the two mediums. While generative design yields one best candidate after multiple interactions, the AR environment results could vary depending on the varying physical landscapes. It is, therefore, we need a layout algorithm to form a bridge between constraints and digital objects.

The current design process heavily relies on flat-screen devices – computers, mobiles, or even conventional pen and paper. Immersive environments like AR are new, offering fresh perspectives for ideation and design exploration. However, these environments also raise challenges in terms of user interfaces and input methods. The absence of tactile interactions like that of a mouse, pen, or a touchpad necessitates AR interfaces to be efficient and user-friendly. Despite being one of the first group proponents of generative design in AR, a scant amount of informational resources like products and studies are available for reference. The ongoing project's user interface involves a detached 2D scene where designers have to step outside of their AR environment to interact with the UI for constraint modifications. To counter this, we suggest a 3D user interface for generative design software in AR, a system integrated into the AR scene offering real-time projection of current constraints using feed-forwarding technology. This system enables adding and deleting constraints without changing scenes and prompt feedback, unlike the traditional constraint list-dependent method. Our proposed UI provides real-time constraint projection via feed-forwarding technology, seeking to enhance user experience and design process in AR. Figure 1 shows one scenario when using the UI system.

## 2. Related Works

Generative design, predominantly utilized in mechanical engineering applications, has witnessed significant advancements over the past few years. Meanwhile, the implementation of generative design within immersive environments, such as augmented reality (AR), is an emerging trend that has started to gain attention. This section reviews the evolution and current state of generative design, as well as AR technology, providing background relevant to our work in developing a UI for generative design software within AR.

### 2.1 Generative Design

Generative design is a burgeoning field that has experienced rapid evolution of late[4,10,13]. It involves a design system led by algorithmic processes, accepting high-level objectives and constraints from designers, and producing an array of alternative designs[21-23]. This approach has found utility in various fields, including architecture and mechanical engineering[3,8,12,20]. Several studies have proposed innovative frameworks for conceptual building design and introduced computational methods to generate organic interface structures for supporting objects in given physical environments[8,11,24]. These works, like ours, illustrate a standard generative design workflow wherein users specify design constraints and explore alternative designs[8,17].

The exploration aspect of generative design, particularly, has benefitted from increasing computational power and advancements in artificial intelligence[14]. With these advancements, the scope of generative design solutions has expanded, prompting extensive research into exploration strategies for these solutions. Several tools have emerged, such as GEM-NI, a graph-based generative design tool that promotes simultaneous exploration of multiple design alternatives, and the "creative optimization tool," which facilitates exploration of potential solutions and analysis of design trade-offs[2,25].
Additionally, many studies are merging different technologies with generative design to enable the exploration of thousands of design alternatives. Tools like Dream Lens, an interactive visual analysis tool, and V-dream, a virtual reality generating analysis framework, depict the immense possibilities of this integration[26,27]. Our work follows this trend, combining AR with generative design to assist designers in zeroing in on the most suitable solution.

### 2.2 Augmented Reality

Augmented Reality (AR) technology has surged in popularity as an innovative tool for incorporating digital assets into a physical environment context[7,28,29]. This technology manifests in several forms including wearables and smart glasses that use retinal projection, such as Google Glass and Vaunt by Intel, as well as in the form of widely used smartphones[6,15]. Originally, AR predominantly operated based on pre-existing understanding of the environment, often using square markers due to their simplicity and effectiveness[5,30]. However, the necessity of a detector marker was a stark limitation, leading to the development of markerless AR, which leverages plane detection, GPS, and object detection technologies[31].

As AR gained popularity and expanded into various fields, there was a growing accent on enhancing user-friendliness and usability[1,9,18]. Context-awareness emerged as one of the methods to achieve this goal[32,33]. It allows AR applications to adapt to the user's situation and needs, rendering the application more ergonomic and user-friendly[32-34]. Context changes, particularly in mobile AR where users move around and use their devices in diverse environments, influenced the industry to develop services that offer customized user experiences[34]. For example, Unity Mars simulates AR in the virtual environment and allows users to alter the environment continuously, resulting in faster workflows for AR development and visually appealing AR apps[19]. Similarly, 360 Proto is a tool that rapidly generates interactive virtual reality and AR prototypes from paper[35].

AR has found several applications, such as previewing products in the user's environment, exploring prototype builds of a product, and more[16,36]. This project uses AR to assist designers in checking their design's functionality and accessibility in a virtual 3D environment. In this context, AR serves as a suitable test environment for real-world applications.

## 3. Method

This section delves into the specifics of our proposed system and its implementation, beginning with a detailed overview of the interface, followed by a thorough elucidation of the design components, and finally, the operations that allow the system to function as intended.

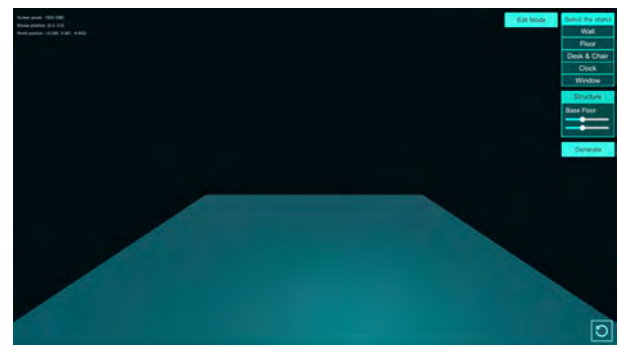

Figure 2 The initial user interface

### 3.1 System Description

In our study, we propose a 3D user interface system for generative design within AR software. Within this interface, users can articulate their design constraints graphically, prompting the system to generate every possible constraint automatically. To demonstrate the system, we selected a "classroom" scenario, utilizing clocks, windows, and desks as our virtual objects. The clocks and windows are designed to be integrated into the vertical plane, while the desks are designed to sit on the horizontal plane. The initial interface on the system's commencement is represented in Figure 2.

This system operates in two modes – normal and edit mode. During normal mode, users can define a constraint by visually demonstrating the relationship between an object and a plane, or between two objects. For instance, to input the constraint "a clock should be placed on a vertical plane", a user can initially construct a wall, then place a clock on this wall. When the user selects the "Generate" button, this constraint will be automatically generated and outlined on a constraint list panel. Meanwhile, during edit mode, users can shift or delete an existing object. The constraints tied to that object will then be automatically updated following the edit action. For instance, if a clock and window reside on the same wall yet are not aligned, and the user desires them to be aligned along the X-axis, the user can initially switch to "move" mode and then arrange the two objects on the same X-axis. The alignment constraint will be added to the constraint list following the move action. Similarly, all constraints linked with one object will be deleted if that object is removed.

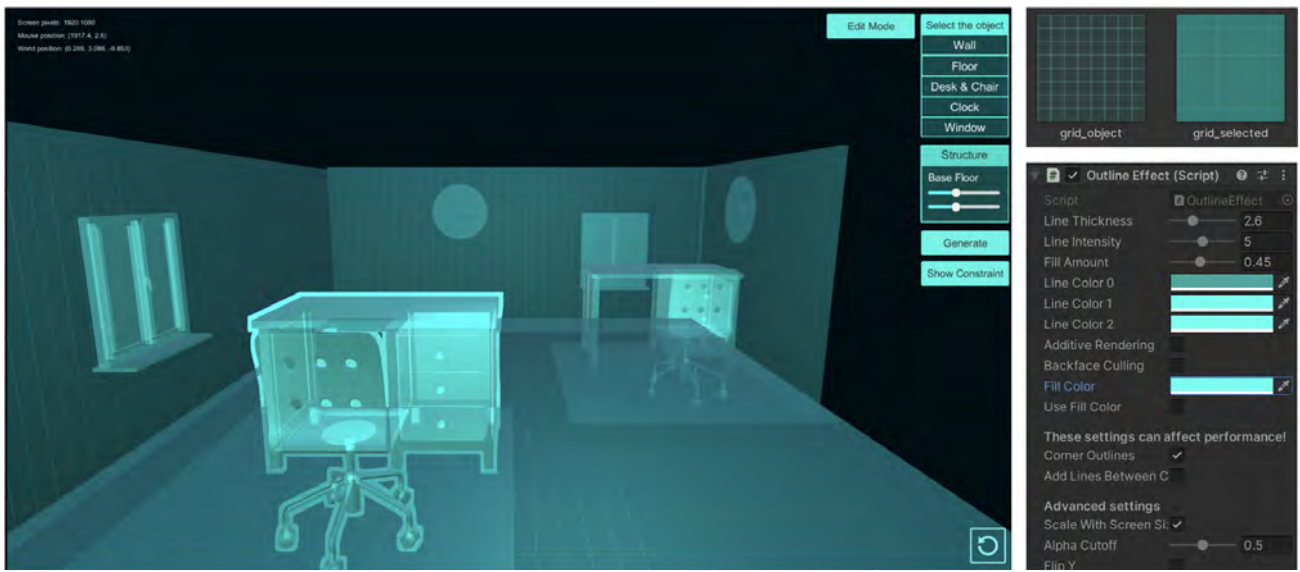

Figure 3 Overall UI system design and graphic component

Users are able to view the current constraints on a specific object or generate the total constraint list at any point during the construction process. They can also edit the objects and planes if they are unsatisfied with the generated constraints. Once satisfied with the constraint list, users can utilize the export function to have the constraint written into a CSV file. This file can then be parsed by the backend system, resulting in the generation of all possible designs in various environments upon constraint analysis.

### 3.2 User Interface

Our front-end user interface was designed with a user-centric perspective, aiming to provide an intuitive and efficient environment for the users to customize their virtual 3D space. The interface encompasses four types of edit modes, an object selection panel, and a structure adjustment panel. Unity's Prefab system was utilized for creating and storing objects and child objects as reusable assets for further efficiency.

The four edit modes include the default, move, scale, and remove. In the default mode, users are facilitated to generate selected objects at any location clicked using the mouse. Furthermore, the objects are automatically oriented to harmoniously fit the space, dependent on whether objects are positioned near a corner, attached to a vertical plane, or attached to a horizontal plane. In 'move' mode, the interface allows users to shift selected objects from their original positions to new positions, using the same embedded algorithm as the default mode. For added flexibility, in 'scale' mode, users can choose to enlarge or reduce the size of the selected objects as per their need for smaller or larger virtual space. This functionality is complemented by a sliding bar that appears, aimed to conveniently adjust the object scale. In the 'remove' mode, users can delete the selected objects, thus nullifying their corresponding constraints if undesired. A pop-up window is generated for confirmation when this function is selected to avoid accidental deletions.

The object selection panel encompasses five objects – walls, floors, desks and chairs, clocks, and windows – categorized into 'structures' and 'items'. Users can build the structure of the 3D space using walls and floors, while items can be attached to these structures to generate desired constraints.

The atmosphere of our project is set as a semi-transparent blue neon space to foster an immersive virtual environment. Initially, we incorporated various textures for the objects. However, it was observed that these textures influenced user perception and encumbered their imagination while customizing the space. Hence, the decision was made to keep the space neutral with a uniform texture. This approach allows users to shape and envision their unique space without any distraction or implicit guidance from texture aesthetics. Lastly, a light blue color outline and a highlighted texture were implemented for the selected objects, to improve visual recognition for users.

### 3.3 Implementation

We utilized Unity as our development platform, integrating an Input System APIs – an extended official package from Unity. In order to accurately interpret constraints based on object positions, we implemented a 'Constraint' class. This class encompasses all currently available constraint types (as shown in Algorithm 1). To efficiently manage the positions and constraints of objects within the AR space, we have established a global dictionary for objects and planes. The object dictionary links each object to an instance of 'Constraint', enabling seamless tracking and alterations of relationships between different objects. This approach simplifies the process of retrieving current constraints on a specified object. The plane dictionary, on the other hand, maintains a list of all objects placed on each plane. To generate an overarching list of constraints, our system iterates through both dictionaries. Since it is possible for duplicate constraints to exist in the object dictionary, a crucial part of the process is ensuring these duplicates are avoided during the generation. The Algorithm 1 provides the pseudo-code outlining the interactive processes embedded within our system.

ALGORITHM 1: Interactive Process

```
Class Constraint {
        attach_to_vertical_plane: bool
        attach_to_horizontal_plane: bool
        same_vertical_plane_with: list of objects
        same_horizontal_plane_with: list of objects
        align_x_with: list of objects
        align_y_with: list of objects
        align_z_with: list of objects
}
object_dictionary = { object: Constraint }
plane_dictionary = { plane: list of objects }

while true:
        if add a plane:
                add key to plane_dictionary
        if add an object:
                add key to object_dictionary
                add constraint to object_dictionary and plane_dictionary
        if remove an object:
                remove this object's constraints from object_dictionary and plane_dictionary
                remove key from object_dictionary
        if move an object:
                remove object
                add object again
        if generate constraint:
                iterate through object_dictionary and plane_dictionary
                generate constraints without duplication
end
```

## 4. Conclusion

As detailed in this paper, we offer an innovative 3D user interface system catered for generative design, providing a substantially enhanced, graphical design experience. The principal users of this approach are designers who can now visually define design requirements and constraints via the new system interface, therefore streamlining the generative design process. Following user graphical input, an output file containing the constraints among objects and virtual surfaces is automatically generated in CSV format. Our UI system detects the location of relationships among different objects as well as the relationships between objects and surfaces, which significantly improves the designer's experience. While this development is significant, there remains immense scope for continuous improvement, enhancement and innovation within the field of generative design.

Future work may encompass developing new ways of interacting within the 3D space and improving the back-end capabilities, such as expanding the types of constraints, adding custom constraints, and syncing with a constraint solver server to provide real-time output based on generated constraint lists. Our system currently allows for seven types of auto-generated constraints but could potentially incorporate others, like pairwise distance setting, sequential distance setting, and incorporating different levels of preference per constraint. Future advancements in the field could even enable designers to input their own unique and specific constraints via text input, rather than graphical input.

Additionally, we could take further steps to link our current UI system with a live constraint solver, which would take our system beyond simply providing a CSV file of the constraints to offering real-time feedback on constraint appearances in a variety of virtual environments. This will ensure continuous advancement in the realm of generative design.